\documentclass[]{jkas}

\usepackage{graphicx}
\usepackage{amsmath}
\usepackage{hyperref}

\def\year{2014} 
\def\month{March} 
\def\volume{00} 
\def\issue{0} 
\def\beginpage{1} 
\def\endpage{99} 
\def\received{---} 
\def\accepted{---} 
\date{Received \received ; accepted \accepted}

\title{New Non-Parametric Approach\\ to Determine Proper Motion of Star Clusters}
\author[1,2]{Rhorom~Priyatikanto}
\author[1]{Mochamad Ikbal~Arifyanto}

\affil[1]{Astronomy Program, Institut Teknologi Bandung, Indonesia; \email{rho2m@hotmail.com, ikbal@as.itb.ac.id}}
\affil[2]{National Institute of Aeronautics and Space, Indonesia}

\def\corrauthor{R. Priyatikanto}
\def\runningauthor{R. Priyatikanto \& M. I. Arifyanto}
\def\runningtitle{New Non-Parametric Approach}
\def\keywords{proper motion --- open clusters and associations}

\newcommand{\aena}{A\&A}

\newcommand{\aj}{AJ}
\newcommand{\jcgs}{Journ. of Comp. \& Graph. Stat.}

\def\abstracttext{
The bulk motion of star clusters can be determined after careful membership analysis using parametric or non-parametric approaches. This study aims to implement non-parametric membership analysis based on Binned Kernel Density Estimator which accounts measurements errors (simply called BKDE-e) and to determine the average proper motion of each cluster. This method is applied to 178 selected star clusters with angular diameter less than 20 arc minutes. Proper motion data from UCAC4 are used for membership determination. Non-parametric analysis using BKDE-e successfully determined the average proper motion of 129 clusters, with good accuracy. Compared to COCD and NCOVOCC, there are 79 clusters with less than $3\sigma$ difference. Moreover, we are able to analyse distribution of the member stars in vector point diagram which is not always normal distribution.}

\begin{document}
\jkashead

\section{Introduction}
In the study of stellar cluster population in the Galactic disk, membership analysis or decontamination process is an important process prior to the physical, astrometric and photometric analysis. This process can be conducted based on the proper motion data through parametric \citep{zhao90} or non-parametric \citep{cano90}. The emergence of non-parametric approach for membership analysis mainly due to the lack of parametric functions to match the actual distribution of proper motion data over vector point diagram (VPD). The distribution does not always conform normal distribution as assumed in parametric approach. Thus, non-parametric density estimation (e.g. Kernel Density Estimation, KDE) serves alternative method to analyse the kinematic distribution of cluster members and finally decontaminate them from field stars.

However, kernel-based non-parametric method usually requires more computational effort ($)(n^2)$), especially in the age of all-sky survey. Binning scheme as discussed by \cite{wand94} may provide a way out. The so-called Binned Kernel Density Estimation \citep[BKDE-e][]{priyatikanto14} is used in this study to decontaminate the star clusters from non-member stars and also to determine the bulk motion of the cluster in the celestial plane. This method takes measurement errors into calculation.

\section{BKDE-e in Brief}
In BKDE-e membership analysis \citep{priyatikanto14}, the proper motion data are binned using linear binning. The number of bins is determined appropriately to minimize error in density estimation \citep{wand94}. After binning, the kernel density estimation is conducted for every knots to model the kinematic distribution of stars within sampling radius ($f_{c+f}$) and the field stars in the annulus ($f_{f}$). Then, the membership probability can be calculated using:
\begin{equation}
P=\dfrac{f_{c+f}-f_f}{f_{c+f}}.
\end{equation}

\begin{figure}[ht]
\centering
\includegraphics[width=0.4\textwidth]{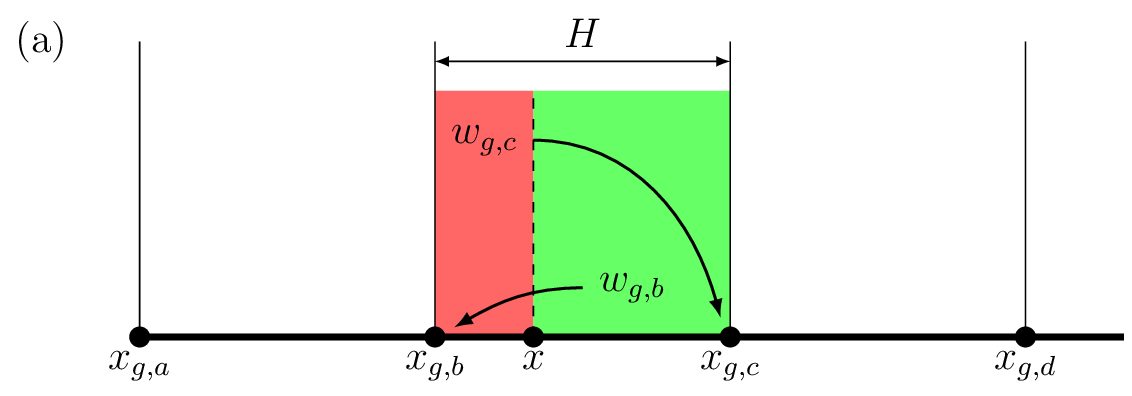}\\
\includegraphics[width=0.4\textwidth]{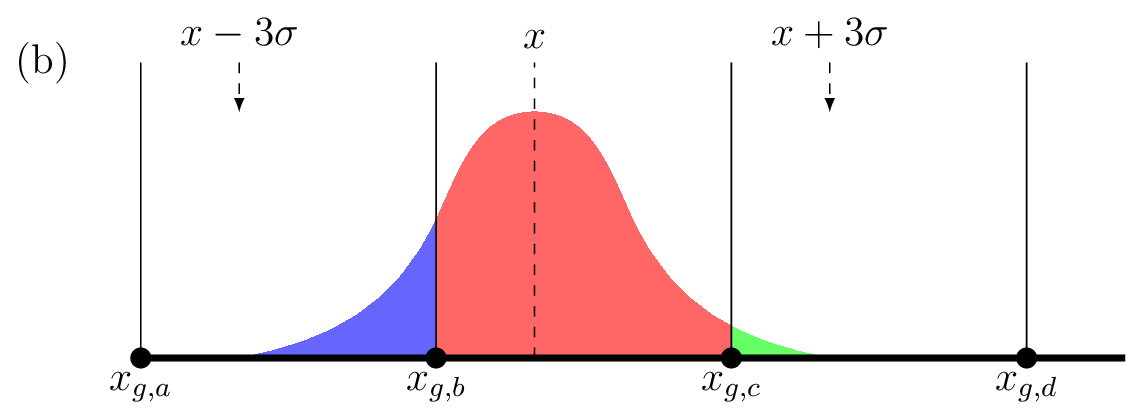}
\caption{Linear binning scheme of $1D$ data point $x\pm\sigma$ contained within $H$-sized bin. Instead of giving its weight to two neighbouring knots ($x_{g,c}$ and $x_{g,c}$) (figure a), data point with error may give its weight to more than two knots enclosed by its $3\sigma$ wings (figure b).}
\label{binning}
\end{figure}

In the BKDE-e ($1D$ case), density estimate for each knot (binning representative, $x_{g,i}$) is calculated using:
\begin{equation}
{f}(x_g)=\dfrac{1}{nH}\sum_{i=1}^{n_{g,x}}K\left(\dfrac{x_g-x_{g,i}}{h_x}\right)c_{g,i},
\end{equation}
where $H$ is kernel width, $K(x)$ is kernel function (Gaussian is used here), while $c_{g,i}$ is the count for each knot that represents the total weights from neighbouring data points. Linear binning scheme is used to obtain the count for each knot (see \autoref{binning}). To accommodate measurement errors, each data point give its partial weight to the related knots according to the total area enclosed by two knots:
\begin{align}
\label{eq:integral1}
w_{g,b}&=\int_{x_b}^{x_c}g(x')dx'-\dfrac{1}{H}\int_{x_b}^{x_c}(x'-x_b)g(x')dx',
\\
\label{eq:integral2}
w_{g,c}&=\dfrac{1}{H}\int_{x_b}^{x_c}(x'-x_b)g(x')dx'.
\end{align}

Bulk motion or average proper motion of the cluster is determined according to the mode of the kinematic distribution on the VPD. Stars with $f_{c}>90\%f_{c,max}$ are averaged to determine the average proper motion. This approach is more robust compared to the average proper motion of member stars because of the asymmetrical distribution.

\section{Kinematic Data}
The Fourth US Naval Observatory CCD Astrograph Catalog \citep[UCAC4][]{zacharias13} that includes $\sim105$ millions stars with proper motion data and typical error less than 10 mas/yr is as the source of kinematic data in this study. We select 178 open clusters with radius $R<1^{\circ}$ and $1\sigma$ members $N_1>10$ as catalogued by \cite{kharchenko05}. Stars with photometric error $e_J<0.1$ and kinematic error $\mu<7$ mas/yr are used in membership analysis. The distance of selected clusters range from 0.3 to 6.0 kpc, while the age range is $\sim4$ Myr to $\sim2$ Gyr.

Among the selected stars of every cluster, we categorize them into \emph{in-field} and \emph{out-field} stars. In-field stars ($r<R_{tide}$) consist of both member and field stars, while out-field stars ($1.2R_{tide}<r<1.6R_{tide}$) contain only field stars.

\section{Result and Discussion}
\begin{figure}[ht]
\centering
\includegraphics[height=3.5cm,trim=0.2cm 0 8.5cm 0,clip]{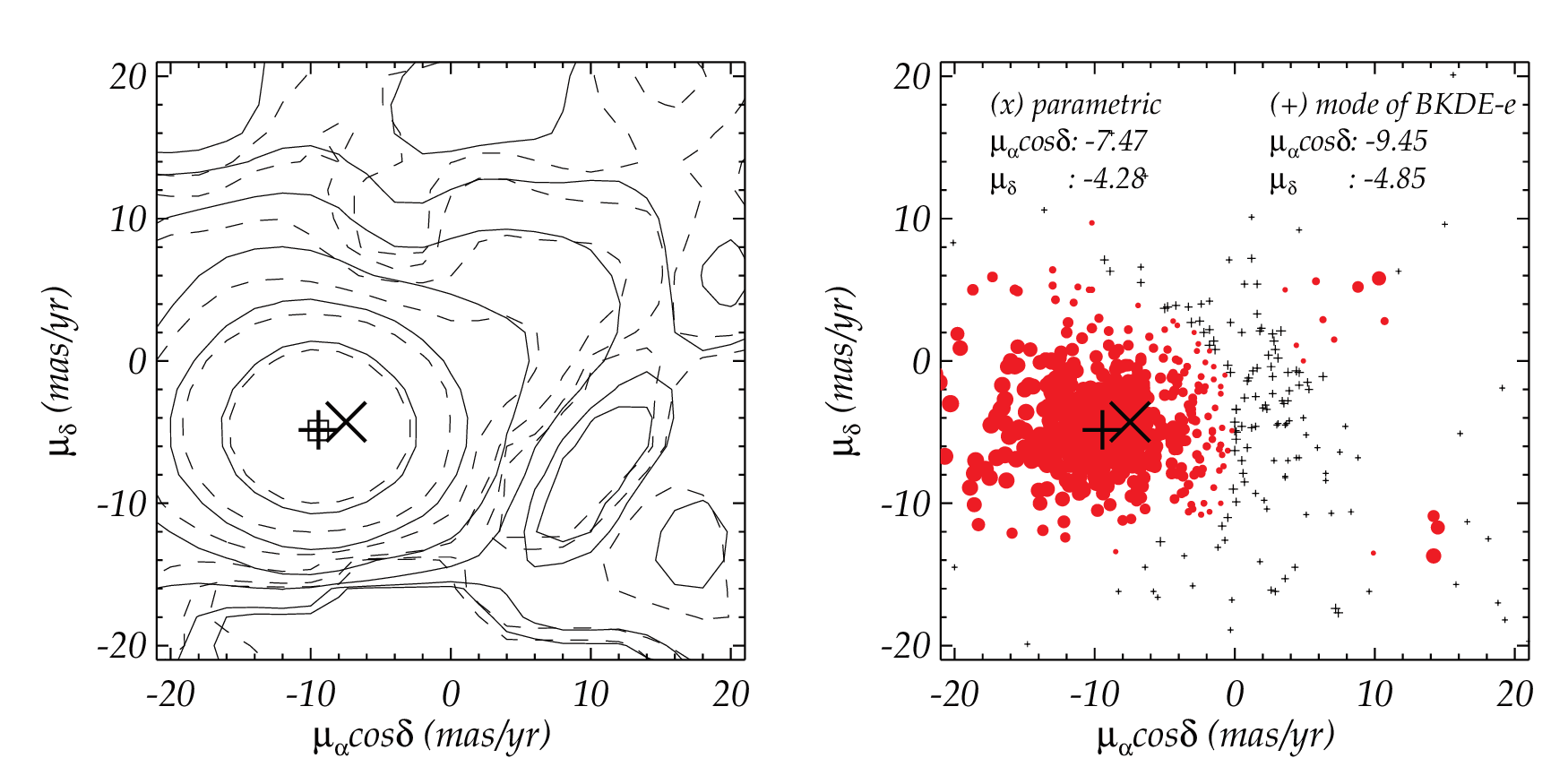}
\includegraphics[height=3.5cm,clip,trim=8cm 0.2cm 0.2cm 0.2cm]{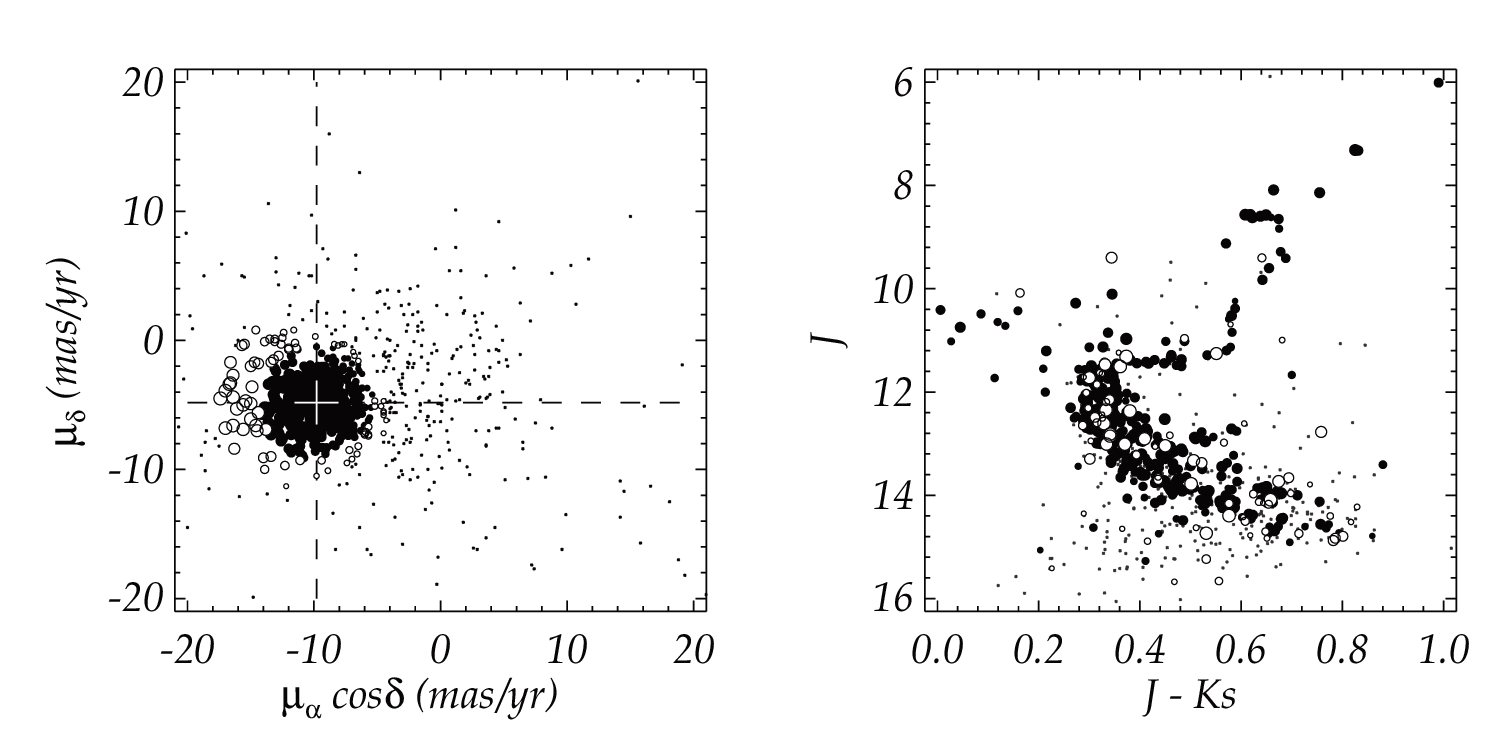}
\captionof{figure}{\emph{Left}: Contour of density estimate (on VPD) of stars around NGC 2682 obtained using BKDE-e (solid) and ordinary KDE (dashed), plus marks the average proper motion obtained using BKDE-e. \emph{Right}: the location of cluster members (filled circles) and non-members (open circles) over the CMD.}
\label{ngc2682}
\end{figure}

Kinematic analysis was conducted to 178 selected clusters. Among those samples, we successfully obtained proper motion of 129 (72\%) with median uncertainty of 1.5 mas/yr. One example of the successful cases is NGC 2682 as displayed in \autoref{ngc2682} The other 49 (28\%) clusters have low concentration or embedded within the cloud. These condition increase the difficulty in decontamination process by non-parametric approach. In the most of the unsuccessful cases, density estimate of the in-field stars are indifferent compared to the out-field stars.

\begin{figure}[ht]
\centering
\includegraphics[width=0.39\textwidth]{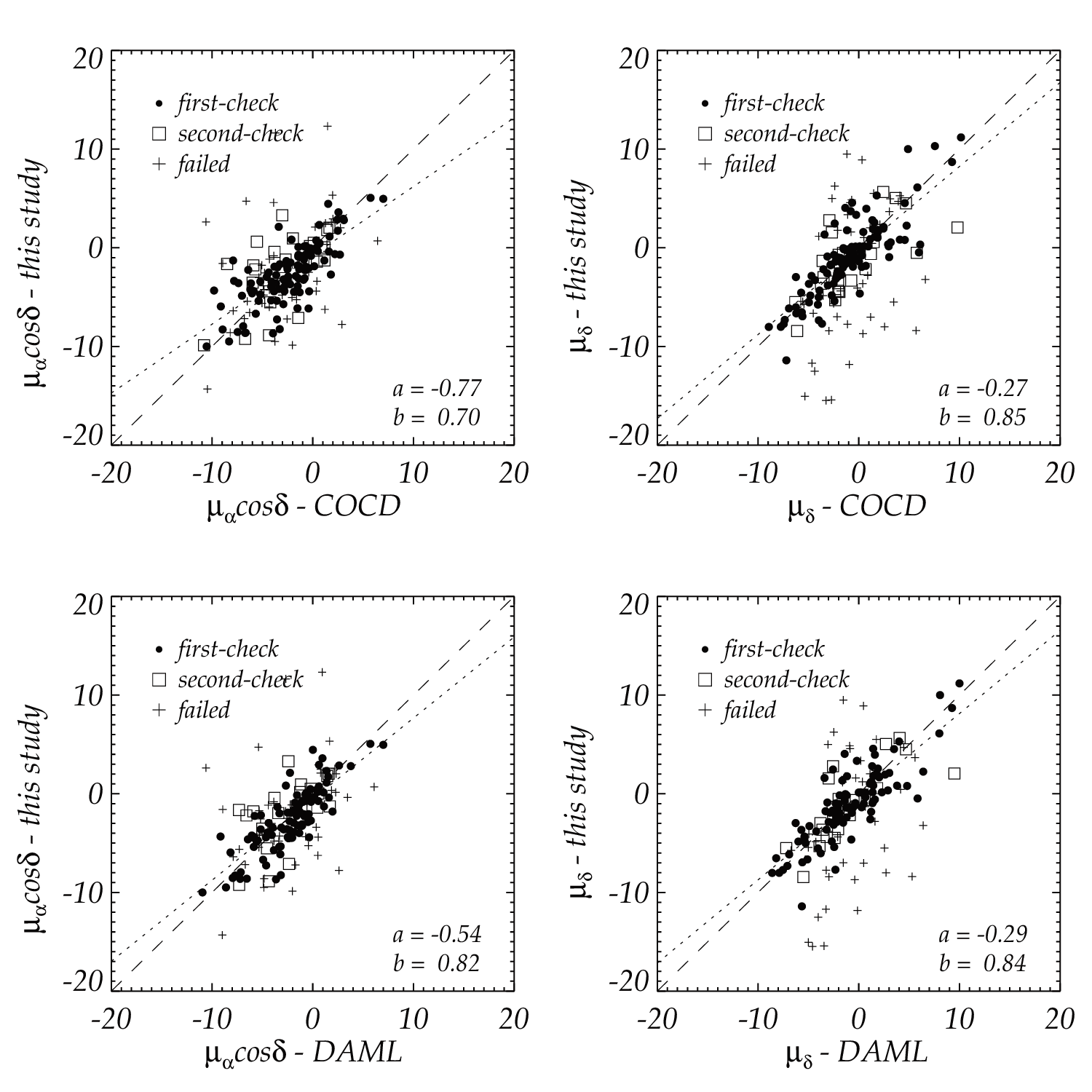}
\caption{Plots of cluster proper motion obtained in this study compared to COCD (top) and NCOVOCC (bottom). Dotted lines mark the linear fit of the data that show a small deviation, fitting coefficients ($a,b$) are displayed altogether.}
\label{compare}
\end{figure}

The obtained results were compared to published proper motion catalog: Catalog of Open Cluster Data \citep[COCD][]{kharchenko05} New Catalog of Optically Visible Open Clusters and Candidates \citep[NCOVOCC][]{dias02} (see \autoref{compare}). For quantitative comparison, we defined:
\begin{equation}
\Delta_{\mu}^2=\dfrac{(\mu_x-\mu_{x,cat})^2}{\sigma_x^2+\sigma_{x,c}^2}
+\dfrac{(\mu_y-\mu_{y,cat})^2}{\sigma_y^2+\sigma_{y,c}^2},
\end{equation}
where $\mu_x=\mu_{\alpha}\cos\delta$ and $\mu_y=\mu_{\delta}$. Among the 129 successful cases, 61\% of our results agree to with catalog ($\Delta_{\mu}<3$).

\vskip-10pt

\acknowledgments
This study is supported by decentralized research grant from Ministry of Education and Culture. RP gratefully acknowledge the travel grant from Faculty of Mathematics and Natural Sciences ITB and from IAU.

\end{document}